\documentclass[preprintnumbers,article,amsmath,amssymb,floatfix,10pt,prd,onecolumn,
superscriptaddress,nofootinbib]{revtex4}
\usepackage[colorlinks=true, pdfstartview=FitV, linkcolor=blue, citecolor=red, urlcolor=magenta]{hyperref}
\usepackage{bbm}
\usepackage{amsfonts}
\usepackage{mathrsfs}
\usepackage{latexsym}
\usepackage{epsfig}
\usepackage{epstopdf}
\usepackage{epstopdf}
\usepackage{graphicx}
\usepackage{amssymb}
\usepackage{amsmath}
\usepackage{dcolumn}
\usepackage{bm}
\usepackage{color}
\usepackage{comment}
\usepackage{xcolor}
\begin{document}

\title{\bf Exploring the viability of charged Spheres admitting non-metricity and matter source}

\author{M. Zeeshan Gul}
\email{mzeeshangul.math@gmail.com}\affiliation{Department of Mathematics and Statistics, The University of Lahore,\\
1-KM Defence Road Lahore-54000, Pakistan}\affiliation{Research
Center of Astrophysics and Cosmology, Khazar University, Baku,
AZ1096, 41 Mehseti Street, Azerbaijan.}

\author{Faisal Javed}
\email{faisaljaved.math@gmail.com}\affiliation{Department of
Physics, Zhejiang Normal University, Jinhua 321004, People Republic
of China}

\author{M. Sharif}
\email{msharif.math@pu.edu.pk}\affiliation{Department of Mathematics and Statistics, The University of Lahore,\\
1-KM Defence Road Lahore-54000, Pakistan}

\author{Shalan Alkarni}
\email{shalkarni@ksu.edu.sa}\affiliation{Department of Mathematics,
College of Sciences, King Saud University, P.O. Box, 2455 Ridyadh
11451, Saudi Arabia}

\begin{abstract}
This research paper investigates the impact of non-metricity and
matter source on the geometry of charged spheres in the presence of
anisotropic matter configuration. We use a specific model of
extended symmetric teleparallel theory to minimize the complexity of
the field equations. Moreover, the feasible non-singular solutions
are used to examine the interior composition of the charged spheres.
The Darmois junction conditions are used to determine the unknown
constants in the metric coefficients. We explore some significant
properties in the interior of compact stars under consideration to
check their viable existence in this modified framework. The
equilibrium state of the charged spheres is discussed using the
Tolman-Oppenheimer-Volkoff equation and stability is analyzed by
sound speed and Herrera cracking approach. We find that the charged
spheres in this theoretical framework are
physically viable and stable.
\end{abstract}

\maketitle

\date{\today}

\section{Introduction}

Stars are luminous spheres of plasma held together by its own
gravity. They are also known as cosmic energy engines which radiate
heat into light, X-rays and ultraviolet rays. Stars sustain
equilibrium by counterbalancing the inner force of gravity with
outer pressure. At star core, nuclear fusion reaction converts
hydrogen into helium and energy in form of light and heat that make
them shine brightly. Scientists gain valuable insights into the
formation and evolution of the stellar objects. Once stars exhaust
their fuel, they may cease to provide enough pressure to resist the
gravitational collapse. Different compact stars depend on their
initial mass formed due to this phenomenon. These compact celestial
objects distinguish themselves from other stars by their lack of
pressure to resist against gravitational collapse. Instead, the
particles in their extremely dense matter adhere to the Pauli
exclusion principle (two identical fermions cannot exist in same
energy level), producing a degeneracy pressure that prevents the
collapsing process of these dense objects. As a result, these
stellar objects have small sizes with stronger gravitational and
magnetic fields compared to normal stars. Among these cosmic
objects, neutron stars are the most captivating objects which serve
a significant role in various astrophysical phenomena. Their
formation and behavior have profound implications to comprehend the
cosmos.

Neutron stars are almost composed of neutrons, hence the name
neutron star. One of the most remarkable features of neutron stars
is their intense magnetic fields. These magnetic fields are
important in various astrophysical processes. Scientists claimed
that the compact stars formed as a result of supernova explosions
\cite{1} and this idea gained support with the discovery of pulsars
(rotating neutron stars) \cite{2}. Researchers have extensively
studied the behavior of pulsars under various conditions to uncover
their fundamental characteristics and understand the underlying
physical mechanisms governing their behavior. Thus, the study of
pulsars offerers profound implications of astrophysics and our deep
comprehension of the cosmos. At very regular intervals, the pulsars
have pulses of radiation that range from milliseconds to seconds.
Pulsars emit beams of electromagnetic radiations out of its magnetic
poles due to strong magnetic fields. This radiation can be observed
only when a beam of emission is pointed towards the earth. Many
researchers examined physical attributes of pulsars under various
considerations \cite{3}-\cite{9}.

Isotropy (equal principal stresses) is a common assumption in the
study of self-gravitating systems, whenever the fluid approximation
is used to describe the matter distribution of the object. This
character of fluids is supported by a large amount of observational
evidence, pointing towards the equality of principal stresses under
a variety of circumstances. However, strong theoretical evidences
presented in the last decades suggest that for certain density
ranges, different kinds of physical phenomena may take place, giving
rise to anisotropy. The number of physical processes giving rise to
deviations from isotropy is quite large in both high and low density
regimes. Thus, for highly dense systems, exotic phase transitions
may occur during the process of gravitational collapse. The
consideration of anisotropic fluids in the context of compact stars
is essential to accurate capture the complex physical phenomena that
govern their structure and stability. The inclusion of anisotropic
pressure allows for a more realistic representation of matter under
extreme conditions, leading to improved theoretical models that
align better with observational data and predictions regarding
stellar evolution, gravitational collapse and oscillation modes. In
the extreme gravitational environments of compact stars, the
pressure and density may not be uniform throughout the star.

Anisotropic fluids account for the directional dependence of
pressure, which can arise from several factors such as nuclear
interactions, rotation and magnetic fields. At high densities, the
interactions between particles can create different pressures in
different directions due to the influence of strong nuclear forces.
Rapid rotation and strong magnetic fields can induce anisotropic
pressure distributions. For instance, a rotating star may experience
greater pressure along its equatorial plane compared to its poles.
Anisotropic models allow for a more flexible and realistic equation
of state that can better describe the thermodynamic properties of
the matter in these stars. In dense astrophysical environments,
matter may undergo phase transitions (e.g., from hadronic matter to
quark-gluon plasma), leading to anisotropic pressure components.
Anisotropic fluids can lead to different stability conditions
compared to isotropic ones. In compact star models, stability
against gravitational collapse and oscillation modes can be better
understood with anisotropic pressures, which might stabilize certain
configurations that would be unstable in isotropic models.
Anisotropic pressure can play a crucial role in avoiding
singularities within the star, allowing for solutions that are
physically viable. The detection of gravitational waves from neutron
star mergers has revealed complex dynamics that can be modeled more
accurately with anisotropic fluid dynamics, reflecting the actual
physical processes at play. Anisotropic models can provide a richer
structure for the field equations, enabling solutions that capture
the complex nature of compact stars. The influence of anisotropy has
been extensively studied in \cite{9a}-\cite{9p}.

In the study of compact stars, the assumption of isotropic pressure
might be convenient for simplification, but it often fails to
capture the complex dynamics present in these objects. The physical
processes inherent in stellar evolution, including dissipative
phenomena and dynamic equilibrium, inherently lead to the
development of pressure anisotropy, reinforcing the necessity to
consider it in our models. Incorporating pressure anisotropy into
the modeling of compact stars is not merely an academic exercise,
but it reflects the underlying physical realities of stellar
dynamics. The inherent dissipative processes during stellar
evolution guarantee that anisotropic pressures will manifest,
shaping the final equilibrium state. Thus, models that overlook this
essential feature risk failing to accurately describe the behavior
of these fascinating astrophysical objects. Herrera \cite{9q}
highlighted an important aspect that the pressure anisotropy is not
merely a consequence of specific physical conditions in compact
objects, but is rather an unavoidable outcome of the stellar
evolution process. The consideration of anisotropic fluids in our
model not only aligns with the physical phenomena expected in
compact objects but also accommodates the natural evolution of
pressure anisotropy during the system history. By adopting
anisotropic fluid models, researchers can better capture the
essential characteristics of compact stars, leading to more accurate
predictions of their behavior such as mass-radius relationships and
stability criteria.

General theory of relativity (GR) proposed by Albert Einstein
revolutionized the concept of gravity. By describing gravity as a
curvature of spacetime due to the presence of mass and energy,
Einstein gravitational theory provides a new perspective on the
fundamental forces in nature. Riemannian geometry relies on a metric
tensor to define distances and angles in a curved space. While
non-Riemannian geometry introduces additional terms such as torsion
and non-metricity to explore more general geometric properties.
Teleparallel theory is one such alternative theory where torsion
represents the gravitational interaction \cite{10}. Weyl proposed
the notion of non-metricity, which specifies the existence of
divergence in the metric tensor. Non-metricity provides the
explanation for the cosmic expansion challenging the dark
energy-based explanation. In symmetric teleparalell theory, the
non-metricity defines the gravitational interaction \cite{11}. Xu et
al \cite{12} generalized the symmetric teleparalell theory by
include the matter content in action, named as $f(\textsc{Q},
\textsc{T})$ theory. There are different forms of modified theories
such as curvature, torsion and non-metricity-based theories
\cite{4a}-\cite{5j}.

Xu et al \cite{15a} examined the diverse avenues of research and the
potential of this theory in reshaping our understanding of cosmic
evolution. Arora and Sahoo \cite{15b} examined that the cosmic
evolution can be effectively described by $f(\textsc{Q},\textsc{T})$
theory. The deep understanding of the early cosmos in this
theoretical framework has been discussed in \cite{15c}. Arora et al
\cite{15d} analyzed that the extended symmetric teleparallel theory
addresses the phenomenon of cosmic acceleration. The matter bounce
scenario in the same background has been studied in \cite{15e},
offering an alternative cosmological scenario to the Big Bang
theory. Godani and Samanta \cite{15f} used several cosmic parameters
to explore the cosmic evolution in the same theoretical framework.
Fajardo \cite{15g} highlighted that this extended theory serves as
alternative to standard cosmic model. Arora et al \cite{15h} used
energy conditions to study the dark universe in
$f(\textsc{Q},\textsc{T})$ theory. Gul et al \cite{15i}-\cite{15m}
examined the geometry of compact spheres with different matter
configuration in the same theoretical framework. Javed and his
collaborators \cite{16a}-\cite{16d} presented various aspects of
anisotropic compact stellar models under the influence of distinct
parameters in different scenarios.

Alternative theories have made great progress based on features of
compact spherical structures. Nashed and Capozziello \cite{23}
examined the interior region of CSOs with the consideration of
anisotropic matter configuration in $f(\textsc{R})$ gravity. Kumar
et al \cite{24} studied the internal dynamics of compact spherical
stellar structures in curvature-matter coupled theory. Dey et al
\cite{25} investigated feasible dense objects using the Finch-Skea
solutions. The geometry of compact spherical structurally with
different matter distribution in the framework of
$f(\textsc{Q},\textsc{T}^{2})$ theory has been studied in
\cite{25a}-\cite{25j}. Majeed et al \cite{26} investigated the
stability of dense spheres candidates using Tolman-Kuchviz solutions
in Rastall theory. The physical properties of anisotropic spheres in
the framework of symmetric teleparallel theory has been investigated
in \cite{27a}-\cite{27c}. The stable spheres with anisotropic matter
distribution in $f(\textsc{R},\phi,\chi)$ theory has been examined
in \cite{28a}-\cite{28c}. Bhar and Pretel \cite{28} conducted a
comprehensive analysis of the anisotropic dense objects in extended
symmetric teleparallel theory. The study of static spherical
structures through observational data has been investigated in
\cite{29}. Rej and Bhar \cite{30} used observational data and
addition correction terms to study the behavior of compact spheres
in $f(\textsc{R},\textsc{T})$ theory. Das et al \cite{31}
investigated anisotropic spherically symmetric structures under
$f(\textsc{R},\textsc{G})$ gravity.

We adopt the following structured approach throughout the paper. In
section \textbf{2}, we provide the necessary theoretical background
to understand the behavior of charged pulsars in this modified
framework. Further, we focus on evaluating unknown parameters by
imposing matching conditions. This process is essential for ensuring
consistency between the interior and exterior solutions of the
charged pulsars. In section \textbf{3}, we explore the viable
features of considered charged spheres through the graphical
behavior of different physical quantities. Additionally, we analyze
the the stability of these stars by sound speed and adiabatic index
methods in section \textbf{4}. Section \textbf{5} serves as a
comprehensive summary of our investigation, highlighting the key
insights gained from studying anisotropic charged pulsars in
$f(\textsc{Q},\textsc{T})$ theory.

\section{$f(\textsc{Q},\textsc{T})$ Theory and Charged Spheres}

The modified action of $f(\textit{Q},\textit{T})$ theory with
electromagnetic field is given by \cite{12}
\begin{equation}\label{1}
\textsc{S}=\frac{1}{2\kappa}\int f(\textsc{Q},\textsc{T})
\sqrt{-g}d^{4}x+\int (\textsc{L}_{m}+\textsc{L}_{e})\sqrt{-g}d^{4}x,
\end{equation}
where
\begin{eqnarray}\label{2}
\textsc{L}_{e}=\frac{-1}{16\pi}\textsc{F}^{\alpha\lambda}\textsc{F}
_{\alpha\lambda}, \quad
\textsc{F}_{\alpha\lambda}=\varphi_{\alpha,\lambda}-\varphi_{\lambda,\alpha}.
\end{eqnarray}
The superpotential is given by
\begin{equation}\label{3}
\textsc{P}^{\gamma}_{~\alpha\lambda}=-\frac{1}{2}\textsc{L}
^{\gamma}_{~\alpha\lambda} +\frac{1}{4}(\textsc{Q}^{\gamma}
-\tilde{\textsc{Q}}^{\gamma})g_{\alpha\lambda}- \frac{1}{4}
\delta^{\gamma} _{~[\alpha \textsc{Q}_{\lambda}]}.
\end{equation}
with
\begin{eqnarray}\label{3a}
\textit{Q}_{c}\equiv \textit{Q}^{~a}_{c~a}, \quad
\tilde{\textit{Q}}_{c}\equiv \textit{Q}^{a}_{~c a}.
\end{eqnarray}
Using Eq.\eqref{3}, we obtain the relation for non-metricity as
\begin{equation}\label{4}
\textsc{Q}=-\textsc{Q}_{\gamma\alpha\lambda}\textsc{P}
^{\gamma\alpha\lambda}=-\frac{1}{4}
(-\textsc{Q}^{\gamma\lambda\zeta}
\textsc{Q}_{\gamma\lambda\zeta}+2\textsc{Q}^{\gamma\lambda\zeta}
\textsc{Q}_{\zeta\gamma\lambda}
-2\textsc{Q}^{\zeta}\tilde{\textsc{Q}}_{\zeta}+\textsc{Q}
^{\zeta}\textsc{Q}_{\zeta}),
\end{equation}
The derivation of this equation is given in \cite{12}. The
corresponding field equations are
\begin{eqnarray}\nonumber
\textsc{T}_{\alpha\lambda}+\textsc{T}^{E}_{\alpha\lambda}&=&
\frac{-2}{\sqrt{-g}} \nabla_{\gamma} (f_{\textsc{Q}}\sqrt{-g}
\textsc{P}^{\gamma}_{~\alpha\lambda})- \frac{1}{2} f
g_{\alpha\lambda} + f_{\textsc{T}} (\textsc{T}_{\alpha\lambda} +
\Theta_{\alpha\lambda})
\\\label{5}
&-&f_{\textsc{Q}} (\textsc{P}_{\alpha\gamma\zeta}
\textsc{Q}_{\lambda}^{~\gamma\zeta}
-2\textsc{Q}^{\gamma\zeta}_{~~\alpha}
\textsc{P}_{\gamma\zeta\lambda}).
\end{eqnarray}
where $f_{\textit{T}}=\frac{\partial f}{\partial \textit{T}}$ and
$f_{\textit{Q}}=\frac{\partial f}{\partial \textit{Q}}$.

To investigate geometry of the compact structures, we consider
\begin{equation}\label{6}
ds^{2}=dt^{2}e^{\xi(r)}- dr^{2}e^{\eta(r)}-d
\theta^{2}r^{2}-d\phi^{2}r^{2}\sin^{2}\theta.
\end{equation}
The matter configuration is considered as
\begin{equation}\label{7}
\textsc{T}_{\alpha\lambda}=\textsc{U}_{\alpha}\textsc{U}_{\lambda}
\varrho+P_{r}\textsc{V}_{\alpha}\textsc{V}_{\lambda}-P_{t}
g_{\alpha\lambda}+ P_{t}(\textsc{U}_{\alpha}\textsc{U}_{\lambda}
-\textsc{V}_{\alpha}\textsc{V}_{\lambda}).
\end{equation}

The Maxwell equations are given by
\begin{equation}\label{8}
\textsc{F}_{[\alpha\lambda;\gamma]}=0, \quad
\textsc{F}^{\alpha\lambda}_{;\lambda}= 4\pi \textsc{J}^{\alpha},
\end{equation}
where $(\textsc{J}^{\alpha}=\sigma \textsc{U}^{\alpha})$ is
four-current and charge density is represented by $\sigma$. The
Maxwell field equation  corresponding to static spherical spacetime
becomes
\begin{eqnarray}\label{10}
\varphi''-\left(\frac{\xi'}{2}+\frac{\eta'}{2}-\frac{2}{r}
\right)\varphi'=4\pi\sigma^{\frac{\xi}{2}+\eta},
\end{eqnarray}
where prime is radial derivative. Integrating Eq.(\ref{10}), we
obtain
\begin{eqnarray}\label{11}
\varphi'=\frac{q(r)}{r^{2}}e^{\frac{\xi+\eta}{2}}, \quad
q(r)=4\pi\int_{0}^{r}\sigma r^{2}e^{\frac{\eta}{2}}dr, \quad
E=\frac{q}{4\pi r^{2}}.
\end{eqnarray}
Here, $q(r)$ is total charge in the compact spheres. Using
Eqs.\eqref{5}-\eqref{11}, we obtain the modified field equations as
\begin{eqnarray}\nonumber
\varrho&=&\frac{1}{2r^{2}e^{\eta}}\bigg[2r\textsc{Q}'(e^{\eta}-1)f_{\textsc{Q}
\textsc{Q}}
+f_{\textsc{Q}}\big(e^{\eta}(2+r\eta'+r\xi')+r\eta'-r\xi'-2\big)
\\\label{12}
&+&fr^{2}e^{\eta}\bigg]-\frac{1}{3}f_{\textsc{T}}(3\varrho+P_{r}+2P_{t})-\frac{q^2}{8\pi
r^{4}},
\\\nonumber
P_{r}&=&\frac{-1}{2r^{2}e^{\eta}}\bigg[2r\textsc{Q}'f_{\textsc{Q}
\textsc{Q}}(e^{\eta}-1)
+f_{\textsc{Q}}\big(e^{\eta}(2+r\xi'+r\eta')-2-r\eta'-3r\xi'\big)
\\\label{13}
&+&fr^{2}e^{\eta}\bigg]+\frac{2}{3}f_{\textsc{T}}(P_{t}-P_{r})+\frac{q^2}{8\pi
r^{4}},
\\\nonumber
P_{t}&=&\frac{-1}{4re^{\eta}}\bigg[-2r\textsc{Q}'\xi'f_{\textsc{Q}
\textsc{Q}}
+f_{\textsc{Q}}\big(2\xi'(e^{\eta}-2)-r\xi'^{2}+\eta'(2e^{\eta}+r\xi')
\\\label{14}
&-&2r\xi''\big)+2fre^{\eta}\bigg]+\frac{1}{3}f_{\textsc{T}}
(P_{r}-P_{t})-\frac{q^2}{8\pi r^{4}}.
\end{eqnarray}

To reduce the complexity of these equations, we consider the
functional form as \cite{12}
\begin{eqnarray}\label{15}
f(\textsc{Q},\textsc{T})=\mu \textsc{Q}+\nu \textsc{T},
\end{eqnarray}
where $\mu$ and $\nu$ are the arbitrary constants. Using
Eqs.\eqref{12}-\eqref{15}, we have
\begin{eqnarray}\nonumber
\varrho&=&\frac{e^{-\eta}}{24\pi
r^{4}(2\nu-1)(\nu+1)}\bigg[e^{\eta}(3(q^{2}(1-2\nu)+8\pi
r^{2}\mu(\nu-1))-(8\pi r^{2}\mu
\\\nonumber
&+&q^{2})\nu)+2\pi r^{2}\mu(12(\nu-1)(r \eta'-1)+3r \nu(\xi'(4-r
\eta'+r \xi')+2r \xi'')
\\\label{16}
&+&\nu(4+2r(-\eta'(2+r \xi')+\xi'(4+r \xi')+2r \xi'')))\bigg],
\\\nonumber
P_{r}&=&\frac{e^{-\eta}}{24\pi
r^{4}(2\nu-1)(\nu+1)}\bigg[e^{\eta}(-3(q^{2}(1-2\nu)+8\pi
r^{2}\mu(\nu-1))+(8\pi r^{2}\mu
\\\nonumber
&+&q^{2})\nu)+2\pi r^{2}\mu(12(\nu-1+r \nu \eta')+3r(\xi'(4
\nu-4+r\nu \eta'-r\nu \xi')-2r
\\\label{17}
&\times&\nu \xi'')+2\nu(r \eta'(2+r \xi')-r(\xi'(4+r \xi')+2r
\xi'')-2))\bigg],
\\\nonumber
P_{t}&=&\frac{e^{-\eta}}{24\pi
r^{4}(2\nu-1)(\nu+1)}\bigg[e^{\eta}(3(q^{2}(1-2\nu)+8\pi r^{2}\mu
\nu)+(q^{2}+8\pi r^{2}\mu)\nu)
\\\nonumber
&+&2\pi r^{2}\mu(2\nu(r \eta'(2+r \xi')-r(\xi'(4+r \xi')+2r
\xi'')-2)+3(r(-(\eta'-\xi')
\\\label{18}
&\times&(r(\nu-1)\xi'-2)+2r(\nu-1)\xi'')-4\nu))\bigg].
\end{eqnarray}

\section{Physical Characteristics of Compact Spheres}

Understanding the behavior and properties of pulsars is crucial to
identify their viable features. Here, we study the viable
characteristics of charged pulsars through graphical analysis. In
studying the impact of various physical parameters on stellar
structures, the graphical representations are often employed to
visualize how these parameters influence the properties and behavior
of stellar objects.

\subsection{Analysis of Metric Coefficients}

To establish the singular free spacetime, the metric functions
($\xi$, $\eta$) must be regular and finite. Therefore, we assume the
non-singular solutions that are regarded as a crucial tool for
determining the precise feasible solutions for interior spacetime,
expressed as \cite{34}
\begin{eqnarray}\label{19}
\xi(r)&=&\ln\bigg[a\big(\frac{r^2}{b}+1\big)\bigg], \\\label{19a}
\eta(r)&=&\ln\bigg[\frac{\frac{2
r^2}{b}+1}{\big(\frac{r^2}{b}+1\big)\big(1-\frac{r^2}{c}\big)}\bigg].
\end{eqnarray}
The unknown parameters can be determined by Darmois junction
conditions. In 1929, the French mathematician Georges Darmois
formulated Darmois conditions, the standards criteria to ensure the
physical and mathematical consistency when matching two different
spacetime solutions at a common boundary, known as a junction
surface. The Darmois conditions are crucial to ensure that the
combined spacetime metric remains a valid solution to the field
equations and a transition across boundary is smooth in a physical
sense. The Darmois conditions are used in a variety of context,
including astrophysical models (where an interior solution must be
matched with an exterior solution), wormholes (connecting two
distinct spacetime regions with a throat in-between) and
cosmological models (involving different phases or regions of the
universe).

We consider the Reissner-Nordstrom spacetime as exterior geometry of
the stars as
\begin{eqnarray}\label{20}
ds^{2}_{+}=\Upsilon dt^{2}-\Upsilon^{-1}d r^{2}-r^{2}(d\theta^{2}
+\sin^{2}\theta d\phi^{2}),
\end{eqnarray}
where
\begin{equation}\nonumber
\Upsilon=1-\frac{2\mathcal{M}}{r}+\frac{\mathcal{Q}^{2}}{r^{2}}.
\end{equation}
The metric coefficients exhibit continuity at the surface boundary
$(r=\mathcal{R})$ as
\begin{eqnarray}\nonumber
g_{tt_+}&=&g_{tt_-}\Rightarrow
a(1+\frac{\mathcal{R}^{2}}{b})=\Upsilon,
\\\nonumber
\quad g_{rr_+}&=&g_{rr_-}\Rightarrow
\frac{1+\frac{2\mathcal{R}^{2}}{b}}{(1-
\frac{\mathcal{R}^{2}}{c})(1+\frac{\mathcal{R}^{2}}{b})}=\Upsilon^{-1},
\\\nonumber
g_{tt,r_+}&=&g_{tt,r_-}\Rightarrow \frac{2a
\mathcal{R}}{b}=\frac{2(\mathcal{M}
\mathcal{R}-\mathcal{Q}^{2})}{\mathcal{R}^{3}}.
\end{eqnarray}
By solving above equations, we get
\begin{eqnarray}\label{21}
a&=&1-\frac{3\mathcal{M}}{\mathcal{R}},
\\\label{22}
b &=&\frac{\mathcal{R}^{3}(\mathcal{R}-3\mathcal{M})}{\mathcal{M}
\mathcal{R}+\mathcal{Q}^{2}},
\\\label{23}
c
&=&-\frac{\mathcal{R}^{4}}{\mathcal{M}\mathcal{R}-2\mathcal{Q}^{2}}.
\end{eqnarray}
Table \textbf{1} contains the observable mass, radius of the stellar
objects and the corresponding constants. We consider
$q(r)=\mathcal{Q}(\frac{r}{\mathcal{R}})^{3}$ to investigate the
impact of charge on the geometry of spheres. Figure \textbf{1} shows
that the behavior of metric elements is positively increasing
throughout the domain which assures that spacetime is non-singular.
\begin{figure}
\epsfig{file=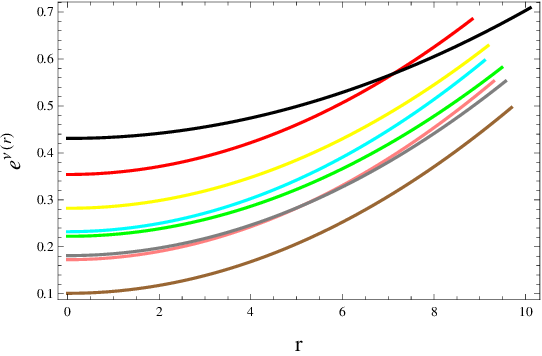,width=.5\linewidth}\epsfig{file=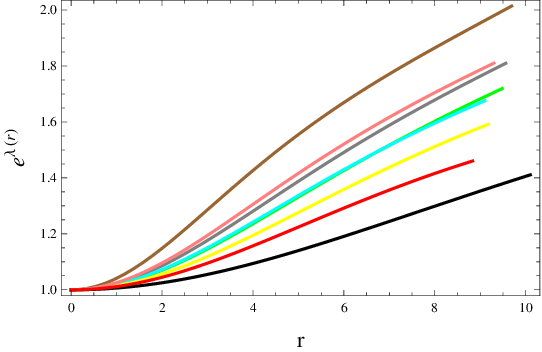,width=.5\linewidth}\caption{Plots
of the behavior of metric coefficients for various charged spheres.}
\end{figure}
\begin{table}\caption{Values of input parameters.}
\begin{center}
\begin{tabular}{|c|c|c|c|c|c|c|}
\hline Compact Stars & $\mathcal{R}(km)$ & $\mathcal{M}_{\odot}$ &
$a$ & $b$ & $c$
\\
\hline EXO 1785-248 [black line] \cite{35} & 1.30 $\pm$ 0.2 & 10.10
$\pm$ 0.44 & 0.430832 & 158.119 & 7687.56
\\
\hline SAX J1808.4-3658 [Blue line] \cite{36}  & 0.9 $\pm$ 0.3 &
7.951 $\pm$ 1.0 & 0.499459 & 102.115 & -536.293
\\
\hline 4U 1820-30 [magenta line] \cite{37}  & 1.58 $\pm$ 0.06 & 9.1
$\pm$ 0.4 & 0.510915 & 102.476 & -483.822
\\
\hline Cen X-3 [yellow line] \cite{38} & 1.49 $\pm$ 0.08 & 9.178
$\pm$ 0.13 & 0.282112 & 68.6541 & 3289.17
\\
\hline SMC X-4 [Red line] \cite{38} & 1.29 $\pm$ 0.05 & 8.831 $\pm$
0.09 & 0.354051 & 83.4878 & -5033.83
\\
\hline Vela X-1 [gray line] \cite{38} & 1.77 $\pm$ 0.08 & 9.56 $\pm$
0.08 & 0.181282 & 44.612 & 1203.25
\\
\hline 4U 1608-52 [pink line] \cite{39} & 1.74 $\pm$ 0.01 & 9.3
$\pm$ 0.10 & 0.172658 & 39.3145 & 1278.23
\\
\hline PSR J1614-2230 [brown line] \cite{40} & 1.97 $\pm$ 0.04 &
9.69 $\pm$ 0.2 & 0.100997 & 23.9767 & 869.679
\\
\hline PSR J1903+327 [green line] \cite{41} & 1.667 $\pm$ 0.021 &
9.48 $\pm$ 0.03 & 0.222418 & 55.6268 & 1525.67
\\
\hline
\end{tabular}
\end{center}
\end{table}

\subsection{Graphical Analysis of Matter Contents}

Fluid parameters including density and pressure (radial and
tangential) are significant for self-gravitating objects. The
behavior of these variables provides valuable relation between
gravitational force and pressure in the pulsars. The dense
composition of these objects leads to the anticipation of fluid
parameters reaching their peak levels at the core of the stars.
Using Eqs.\eqref{14}-\eqref{16}, we have
\begin{figure}
\epsfig{file=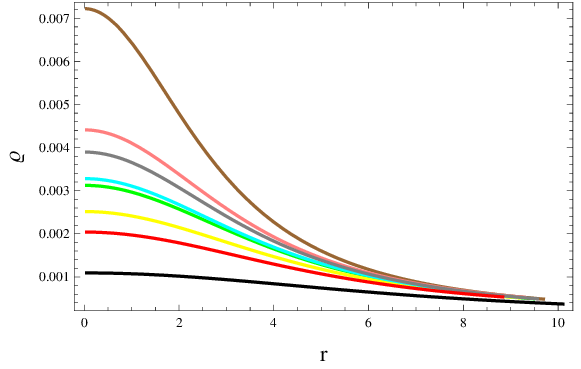,width=.5\linewidth}\epsfig{file=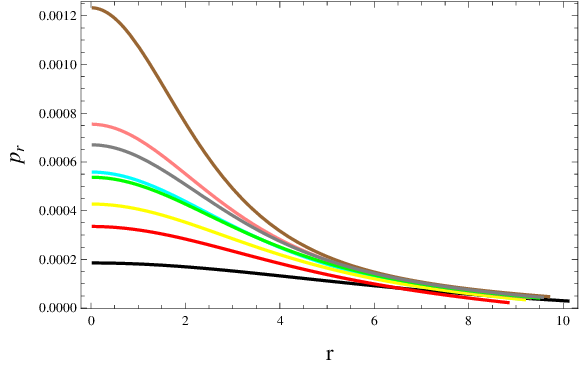,width=.5\linewidth}\center
\epsfig{file=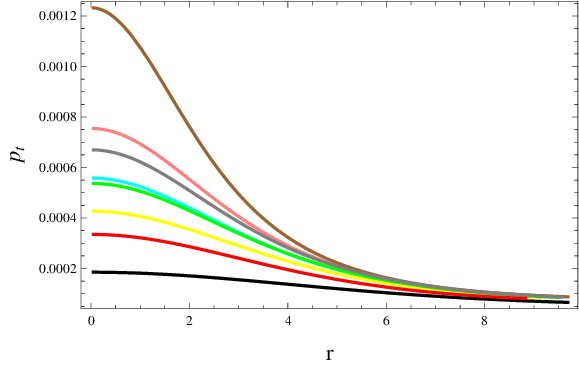,width=.5\linewidth}\caption{Graphical analysis
of matter contents for different charged spheres.}
\end{figure}
\begin{eqnarray}\nonumber
\varrho&=&-\bigg[24 \pi r^{4} (3 b (b + c) + (7 b + 2 c) r^{2} +6
r^{4}) \mu + (7 c q^{2} (b + 2 r^{2})^{2} +24 \pi
\\\nonumber
&\times& r^{4} (2 (b - 4 c) r^{2} +6 r^{4}-b (2 b + 7 c)) \mu) \nu
-3 c q^{2} (b + 2 r^{2})^{2}\bigg]\bigg[ 24 c \pi r^{4}
\\\label{24}
&\times&(b + 2 r^{2})^{2} (1 + \nu) (2 \nu-1)\bigg]^{-1},
\\\nonumber
P_{r}&=&\bigg[3(b + 2 r^{2}) ( 8 \pi r^{4} (b - c + 3 r^{2}) \mu-c
q^{2} (b + 2 r^{2})) + (7 c q^{2} (b + 2 r^2)^{2}
\\\nonumber
&+&24 \pi r^{4} (b (2 b + c) + 6 b r^{2} +6 r^{4}) \mu)
\nu\bigg]\bigg[24 c \pi r^{4} (b +2 r^{2})^{2}(1 + \nu)
\\\label{25}
&\times&(2 \nu-1)\bigg]^{-1},
\\\nonumber
P_{t}&=&\bigg[3(b + 2 r^{2})(c q^{2}(b + 2 r^{2}) + 8 \pi r^{4} (b -
c + 3 r^{2}) \mu) + (-5 c q^{2} (b + 2 r^{2})^{2}
\\\nonumber
&+&24 \pi r^{4} (b (2 b + c) + 6 b r^{2}+ 6 r^{4}) \mu)
\nu\bigg]\bigg[24c \pi r^{4}(b+2r^{2})^{2}(1+\nu)
\\\label{26}
&\times& (2\nu-1)\bigg]^{-1}.
\end{eqnarray}
\begin{figure}
\epsfig{file=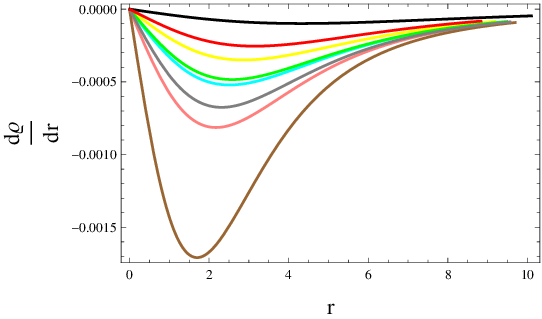,width=.5\linewidth}\epsfig{file=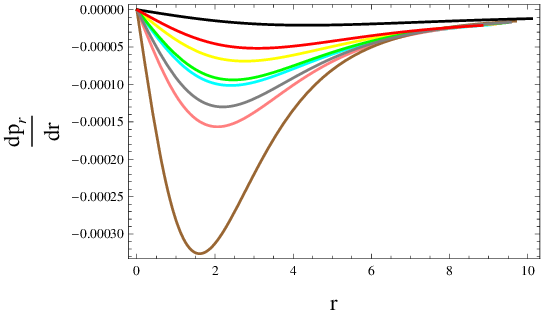,width=.5\linewidth}\center
\epsfig{file=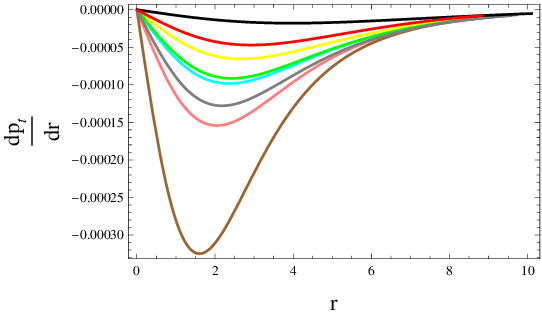,width=.5\linewidth}\caption{Graphs of rate of
change of fluid parameters for various charged spheres.}
\end{figure}
The fluid parameters have their highest value at the center and
thereafter decrease towards the boundary, which provides the dense
description of the proposed stellar objects as shown in Figure
\textbf{2}. Furthermore, the radial pressure in each stellar object
becomes zero at the boundary. Figure \textbf{3} illustrates that the
derivatives of fluid components are zero at the central point of
stellar objects. As distance from the center increases, these
derivatives become negative.

\subsection{Anisotropic Pressure}

Anisotropy describes the variation of pressure across different
directions in a system. Anisotropy holds significance in analyzing
the structural configuration of fluid and their influence on
pressure alignments. When anisotropy is positive, pressure pushes
outward, whereas negative anisotropy leads to inward pressure
\cite{42}. It is usually denoted by $\Delta$, given by
\begin{equation}\nonumber
\Delta=P_{t}-P_{r}.
\end{equation}
The positive anisotropy shown in Figure \textbf{4} evidences the
existence of a repulsive force necessary for cosmic geometries.
\begin{figure}\center
\epsfig{file=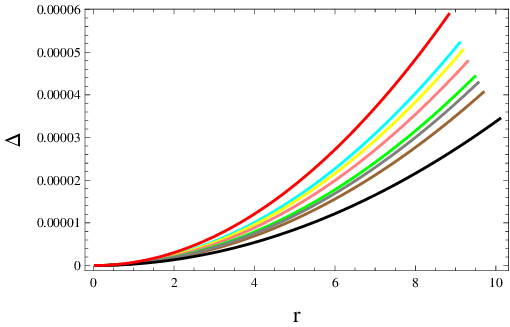,width=.5\linewidth}\caption{Investigation of
anisotropic pressure for different charged spheres.}
\end{figure}

\subsection{Study of Matter in the Pulsars}

Energy conditions are a set of constraints or inequalities that
relate the energy density, flux and pressure in spacetime. These
conditions help to describe the properties of matter and fields that
are consistent with the gravitational theory. The energy conditions
provide restrictions on the EMT required to analyse the feasible
fluid configurations in the system. Here, the following energy
conditions each of which impose different limitations on fluid
parameters as
\begin{itemize}
\item
\textbf{Null Energy Constraint}
\begin{eqnarray}\nonumber
0\leq P_{r}+\varrho, \quad 0\leq P_{t}+\varrho.
\end{eqnarray}
\item
\textbf{Weak Energy Constraint}
\begin{eqnarray}\nonumber
0\leq\varrho, \quad 0\leq\varrho+P_{r}, \quad 0\leq\varrho+P_{t}.
\end{eqnarray}
\item
\textbf{Strong Energy Constraint}
\begin{eqnarray}\nonumber
0\leq\varrho+P_{r}, \quad 0\leq\varrho+P_{t}, \quad
0\leq\varrho+P_{r}+2P_{t}.
\end{eqnarray}
\item
\textbf{Dominant Energy Constraint}
\begin{eqnarray}\nonumber
0\leq\varrho\pm P_{r}, \quad 0\leq\varrho\pm P_{t}.
\end{eqnarray}
\end{itemize}
By considering these energy conditions and their implications on the
energy-momentum tensor,astrophysicists can acquire knowledge about
the characteristics and action of cosmic formations, enhancing our
comprehension of the dynamics and development of the cosmos. Figure
\textbf{5} illustrates that the matter in the charged pulsars is
ordinary as all energy bounds are satisfied when modified terms are
present.
\begin{figure}
\epsfig{file=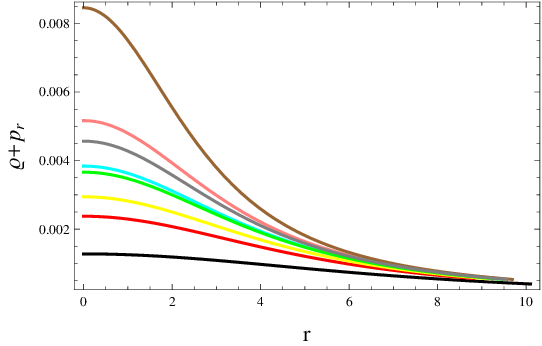,width=.5\linewidth}\epsfig{file=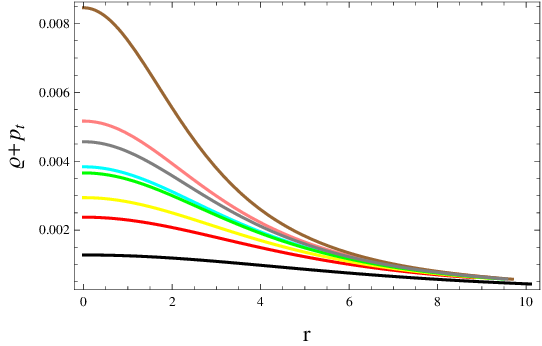,width=.5\linewidth}
\epsfig{file=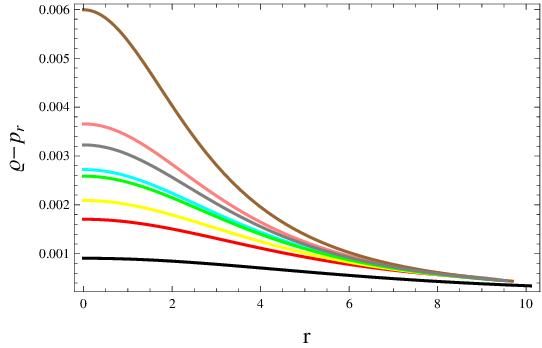,width=.5\linewidth}\epsfig{file=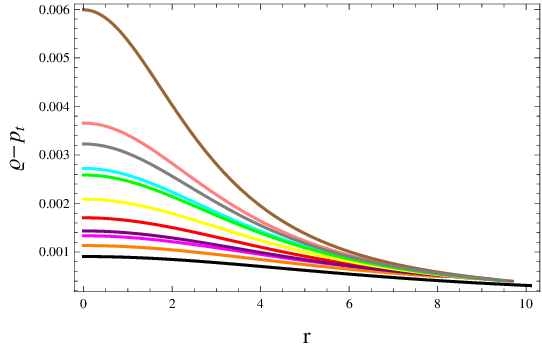,width=.5\linewidth}\caption{Examination
of energy bounds for different charged spheres.}
\end{figure}

\subsection{Relationship between Density and Pressure}

The parameters of the EoS are significant to characterize the
correlation between pressure and density. For viable stellar
objects, the radial component of EoS parameter
$(\omega_{r}=\frac{p_{r}}{\varrho})$ and transverse component
$(\omega_{t}=\frac{p_{t}}{\varrho})$ must satisfy the limit
$(0<\omega_{r}, \omega_{t}<1)$. Using Eqs.\eqref{24}-\eqref{26}, we
have
\begin{eqnarray}\nonumber
\omega_{r}&=&-\bigg[3 (b + 2 r^{2}) (8 \pi r^{4} (b - c + 3 r^{2})
\mu-c q^2 (b + 2 r^{2})) + (7 c q^{2} (b + 2 r^{2})^{2}
\\\nonumber
&+&24 \pi r^{4} (b (2 b + c) + 6 b r^{2} +6 r^{4}) \mu)
\nu\bigg]\bigg[24 \pi r^{4} (3 b (b + c) + (7 b + 2 c) r^{2}
\\\nonumber
&+& 6 r^{4}) \mu + (7 c q^{2} (b + 2 r^{2})^{2} +24 \pi r^{4} (-b (2
b + 7 c) + 2 (b - 4 c) r^{2} +6 r^{4})
\\\nonumber
&\times& \mu) \nu-3 c q^{2} (b + 2 r^{2})^{2} \bigg]^{-1},
\\\nonumber
\omega_{t}&=&-\bigg[3 (b + 2 r^{2}) (c q^{2} (b + 2 r^{2}) + 8 \pi
r^{4} (b - c + 3 r^{2}) \mu) + (24 \pi r^{4} (b (2 b
\\\nonumber
&+& c) + 6 b r^{2} +6 r^{4}) \mu-5 c q^{2} (b +2 r^{2})^{2})
\nu\bigg]\bigg[24 \pi r^{4} (3 b (b + c) + (7 b + 2
\\\nonumber
&\times&c) r^{2} +6 r^{4}) \mu + (7 c q^{2} (b + 2 r^{2})^{2} +24
\pi r^{4} ( 2 (b - 4 c) r^{2} +  6 r^{4}-b (2 b
\\\nonumber
&+& 7 c)) \mu) \nu -3 c q^{2} (b + 2 r^{2})^{2} \bigg]^{-1},
\end{eqnarray}
Figure \textbf{6} shows that the charged pulsars under
considerations are viable as the required limit is satisfied.
\begin{figure}
\epsfig{file=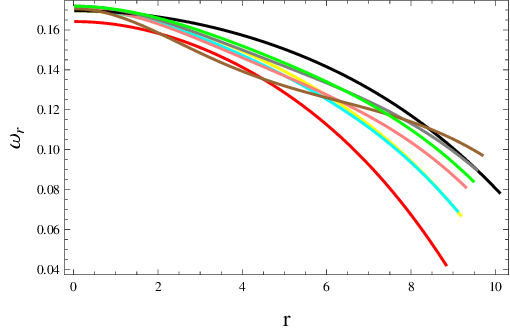,width=.5\linewidth}\epsfig{file=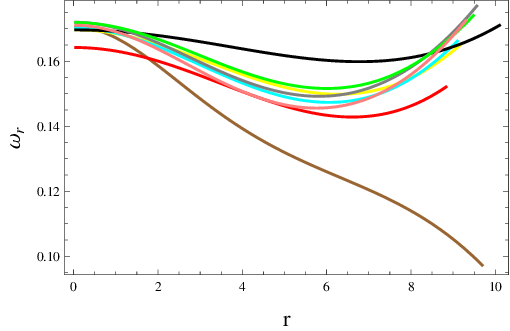,width=.5\linewidth}\caption{Graphical
study of EoS parameters.}
\end{figure}

\subsection{Study of Various Physical Factors}

We use the formula for mass as
\begin{equation}\label{44}
M=4\pi\int^{\mathcal{R}}_{0} r^{2}\varrho dr.
\end{equation}
Figure \textbf{7} determines that the mass function steadily
increase with an increase in radius and  $M\rightarrow 0$ as
$r\rightarrow 0$ at center. This implies that there are no
singularities in the mass distribution at the center. Several
physical aspects can be used to explore the composition of celestial
objects. The compactness function $(u=\frac{M}{r})$ defines a
relation of mass and radius in stellar objects. Buchdahl \cite{44}
assume that compactness factor should be less than 4/9 for viable
spheres. An essential parameter used to comprehend the
characteristics of stellar objects is surface redshift, expressed as
\begin{equation}\label{45}
Z_s =- 1+\frac{1}{\sqrt{1-2u}}.
\end{equation}
For physically viable stellar objects with perfect matter
distribution, Buchdahl \cite{44} proposed that the surface redshift
should be constrained as $(Z<2)$. But, Ivanov \cite{45} found a
higher limit as $(Z<5.211)$ for anisotropic configurations.
Graphical representation of compactness and redshift is shown in
Figure \textbf{8} which indicates that both parameters are
monotonically increasing and satisfied the required viable
conditions.
\begin{figure}\center
\epsfig{file=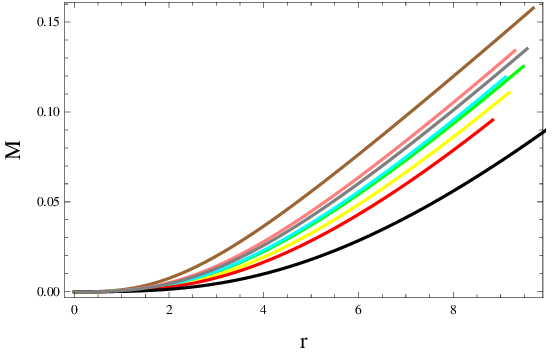,width=.5\linewidth}\caption{Investigation the
behavior of mass function for various charged spheres.}
\end{figure}
\begin{figure}
\epsfig{file=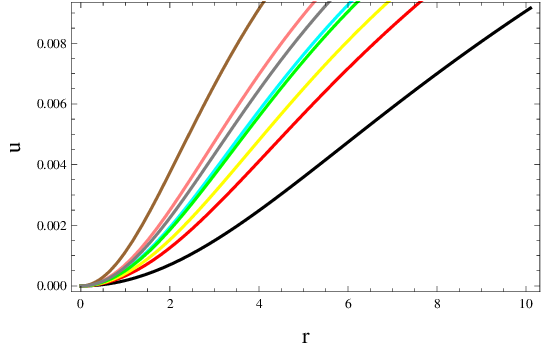,width=.5\linewidth}\epsfig{file=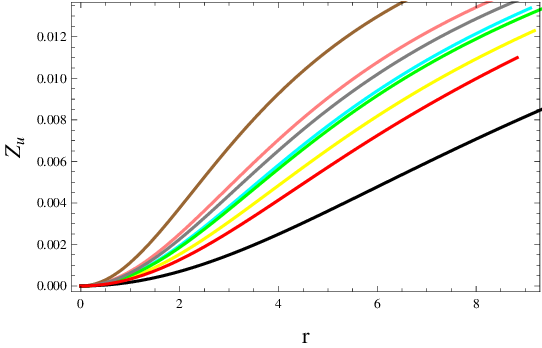,width=.5\linewidth}\caption{Examination
of graphical behavior of compactness and redshift functions.}
\end{figure}

\subsection{Herrera Cracking Condition}

The concept of cracking is associated to the tendency of a fluid
distribution to ``split'', once it abandons the equilibrium as
consequence of perturbations. Thus, one can say that once the system
has abandoned the equilibrium, there is a cracking, whenever its
inner part tends to collapse whereas its outer part tends to expand.
The cracking takes place at the surface separating the two regions.
When the inner part tends to expand and the outer one tends to
collapse we say that there is an overturning. It is worthwhile to
mention that the concepts of stability and cracking are different,
although they are often confused. The term stability refers to the
capacity of a given fluid distribution to return to equilibrium once
it has been removed from it. The fact that the speeds of sound are
not superluminal does not assure in any way the stability of the
object, it only ensures causality. The cracking only implies the
tendency of the system to split immediately after leaving the
equilibrium. Whatever happens next, whether the system enters into a
dynamic regime, or returns to equilibrium, is independent of the
concept of cracking. Of course the occurrence of cracking will
affect the future of the fluid configuration in either case. Herrera
and Di Prisco \cite{45a} developed a general formalism to describe
the occurrence of cracking within a dissipative fluid distribution,
in comoving coordinates.

According to the Herrera cracking method \cite{45b}, a stable
celestial structures exist when the difference of $u_{st}$ and
$u_{st}$ lie within the range of 0 to 1. If the difference exceeds
this range, it implies that the structure is not stable and may be
prone to cracking or collapse. This method helps researchers to
evaluate the structural integrity of the compact structures and
understand the conditions under which it can maintain its stability.
Figure \textbf{9} shows
 the existence of stable charged stellar objects in this theoretical framework.
\begin{figure}\center
\epsfig{file=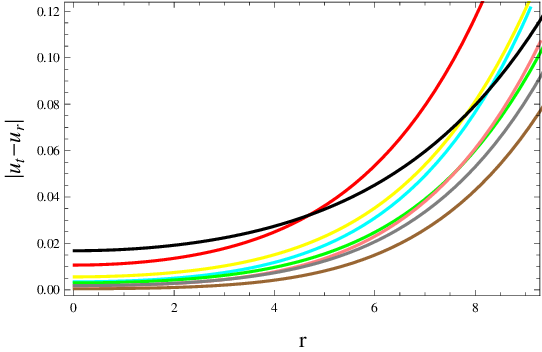,width=.5\linewidth}\caption{Investigation of
cracking for different charged spheres.}
\end{figure}

\section{Equilibrium and Stability Analysis}

Equilibrium and stability is a fundamental factor that holds great
significance to comprehend the behavior of self-gravitating objects.
This analysis examines the behavior of matter in the charged pulsars
and evaluates its capacity to preserve structural integrity and
prevent the collapse. This investigation is significant to
understand the validity and consistency of cosmic structures.

\subsection{Equilibrium State}

The TOV equation describes an equilibrium state of static spherical
symmetric objects \cite{46}, defined as
\begin{equation}\label{46}
\frac{M_{G}(r)e^\frac{\xi-\eta}{2}}{r^{2}}(\varrho+P_{r})+
\frac{dP_{r}}{dr}-\frac{2}{r}(P_{t}-P_{r})=0,
\end{equation}
where
\begin{equation}\nonumber
M_{G}(r)=4\pi \int e^{\frac{\xi+\eta}{2}}(\textsc{T}^{0}_{0} -
\textsc{T}^{1}_{1}-\textsc{T}^{2}_{2}-\textsc{T}^{3}_{3}) r^{2}dr.
\end{equation}
Its solution yields
\begin{equation}\nonumber
M_{G}(r)=\frac{1}{2}r^{2} e^{\frac{\eta-\xi}{2}}\xi'.
\end{equation}
Using Eq.(\ref{46}), we have
\begin{equation}\nonumber
\frac{1}{2}\xi'(\varrho+P_{r})+P_{r}'-\frac{2\Delta}{r}=0.
\end{equation}
This  provides a theoretical framework for comprehending the
internal composition of compact stars and demonstrates the impact of
various forces on the system as
\begin{eqnarray}\nonumber
F_{g}&=&\frac{\xi'(\varrho+P_{r})}{2},
\\\nonumber
F_{h}&=&\frac{dP_{r}}{dr},
\\\nonumber
F_{a}&=&\frac{2(P_{r}-P_{t})}{r}.
\end{eqnarray}
Here, gravitational, hydrostatic and anisotropic forces acting on
the system are represented by $F_{g}$, $F_{h}$ and $F_{a}$,
respectively. Using the field equations \eqref{24}-\eqref{26}, we
obtain
\begin{eqnarray}\label{30}
F_{g}&=&\frac{2 (b + 2 c) r \mu}{c (b + 2 r^{2})^{2} (1 + \nu)},
\\\nonumber
F_{a}&=&\frac{q^{2}}{2 \pi r^{5} + 2 \pi r^{5}\nu},
\\\nonumber
F_{h}&=&\frac{1}{6 c \pi r^{5} (b +2 r^{2})^{3} (1 + \nu) (2
\nu-1)}\bigg[3 c q^{2} (b + 2 r^{2})^{3}+12 (b + 2 c)
\\\nonumber
&\times&\pi r^{6}(b +  2 r^{2}) \mu - (7 c q^{2} (b + 2 r^{2})^{3}
+24 b (b + 2 c) \pi r^{6} \mu) \nu\bigg].
\end{eqnarray}
Figure \textbf{10} shows that the charged spherical objects are in
equilibrium phase as the resultant of all forces is zero.
\begin{figure}\center
\epsfig{file=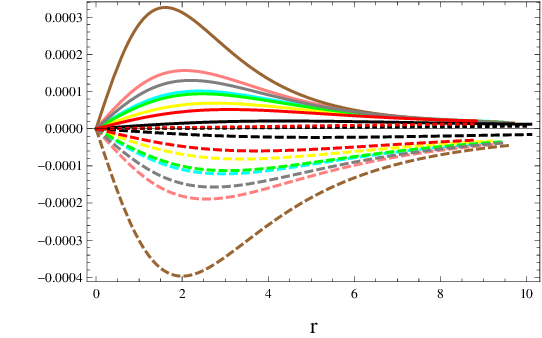,width=.5\linewidth}\caption{Investigation of
different forces $F_{g}$ (dashed line), $F_{h}$ (solid line) and
$F_{a}$ (doted line) on the charged sphere.}
\end{figure}

\subsection{Causality Condition}

The causality principle can be considered which states that no
signal or information can not exceed the speed of light. The sound
speed components are given by
\begin{equation}\label{Ba}
u_{r}=\frac{dP_{r}}{d\varrho}, \quad u_{t} =\frac{dP_{t}}{d\varrho}.
\end{equation}
Using the field equations \eqref{24}-\eqref{26}, we have
\begin{eqnarray}\nonumber
u_{r}&=&\bigg[3 c q^{2} (b + 2 r^{2})^{3} + 12 (b + 2 c) \pi r^{6}
(b +2 r^{2}) \mu - (7 c q^{2} (b + 2 r^{2})^{3}+24 b
\\\nonumber
&\times&(b + 2 c) \pi r^{6} \mu) \nu\bigg]\bigg[12 (b + 2 c) \pi
r^{6} (5 b + 2 r^{2}) \mu + (7 c q^{2} (b + 2 r^{2})^{3} -24
\\\nonumber
&\times&(b + 2 c) \pi r^{6} (5 b + 4 r^{2}) \mu)-3 c q^{2} (b +2
r^{2})^{3} \nu\bigg],
\\\nonumber
u_{t}&=&\bigg[ 12 (b + 2 c) \pi r^{6} (b +2 r^{2}) \mu + (5 c q^{2}
(b + 2 r^{2})^{3} -24 b (b + 2 c) \pi r^{6} \mu) \nu
\\\nonumber
&-&3 c q^{2} (b + 2 r^{2})^{3} \bigg]\bigg[12 (b + 2 c) \pi r^{6} (5
b +2 r^{2}) \mu + (7 c q^{2} (b + 2 r^{2})^{3} - 24
\\\nonumber
&\times&(b + 2 c) \pi r^{6} (5 b + 4 r^{2}) \mu) \nu-3 c q^{2} (b +2
r^{2})^{3} \bigg]^{-1}.
\end{eqnarray}
According to Abreu \cite{47}, the components of sound speed must lie
in [0,1] interval for stable stellar objects. Figure \textbf{11}
shows that the stellar structures under consideration are stable as
they fulfill necessary condition.
\begin{figure}
\epsfig{file=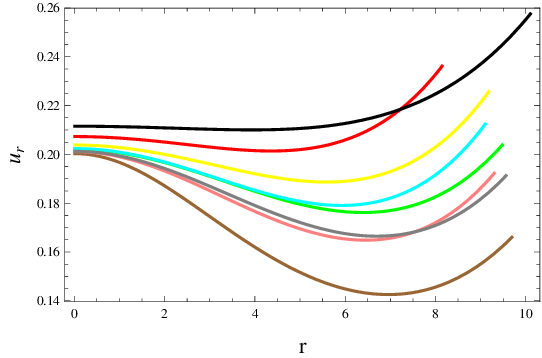,width=.5\linewidth}\epsfig{file=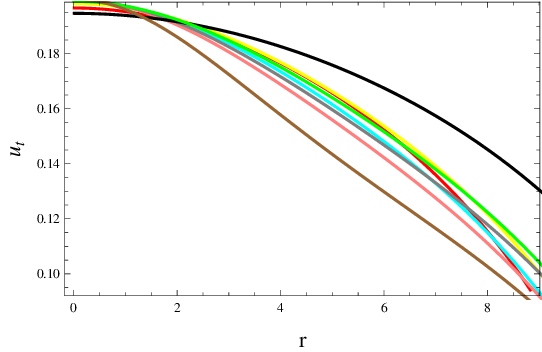,width=.5\linewidth}\caption{Study
of behavior of sound speed components for different charged
spheres.}
\end{figure}

\subsection{Adiabatic Index}

The adiabatic index method is an essential tool in studying the
properties and stability of compact stars. It provides a way to
analyze how a star responds to small perturbations in its structure,
helping to understand the stability of the star. Through this
method, astrophysicists can probe the behavior of matter at extreme
densities, offering insights into the nature of compact objects This
method is instrumental in examining the stability of celestial
bodies and gaining insights into the characteristics of the
substances. Ensuring the stability of these entities is crucial as
any disturbance could lead to collapse or explosion. The adiabatic
index offers insights into how matter reacts to alterations in
pressure and density, aiding in the determination of the stability
of a compact star. The components of the adiabatic index are defined
as
\begin{eqnarray}\nonumber
\Gamma_{r}=\frac{\varrho+P_{r}}{P_{r}}\frac{dP_{r}}{d\varrho},\quad
\Gamma_{t}=\frac{\varrho+P_{t}}{P_{t}}\frac{dP_{t}}{d\varrho},
\end{eqnarray}
where, $\Gamma_{r}$ and $\Gamma_{t}$ are the radial and tangential
components of adiabatic index. To check the stability of a compact
star using the adiabatic index method, one needs to calculate the
value of $\Gamma$ \cite{48}-\cite{51}. If the value of $\Gamma$ is
less than 4/3 then the compact star is stable. If the value of
$\Gamma$ is greater than 4/3, the compact stars is unstable and will
collapse. This means that small perturbations can cause the material
in the star to mix and lead to a loss of energy. As a result, the
star can become less stable and may eventually collapse. Figure
\textbf{12} shows that our system is stable in the presence of
higher-order curvature and matter source terms. Hence, we obtain
viable and stable compact stars in this modified framework.
\begin{figure}
\epsfig{file=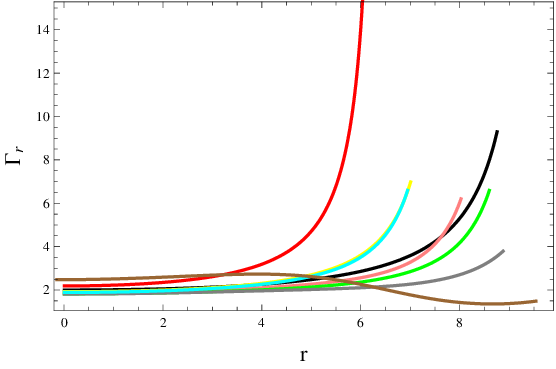,width=.5\linewidth}\epsfig{file=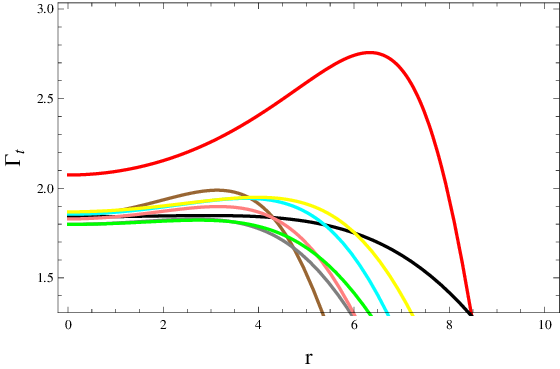,width=.5\linewidth}\caption{Graphical
behavior of adiabatic index for different charged spheres.}
\end{figure}

\section{Conclusions and Discussion}

This study delves into examining the feasibility and stability of
charged pulsars in extended symmetric teleparallel theory.
Specifically, we aim to explore whether the inclusion of
non-metricity and the trace of the energy-momentum tensor in the
gravitational field equations leads to viable solutions for charged
pulsars. Furthermore, we have analyzed that graphical analysis of
different physical properties to ensure the viability of the charged
pulsars in the proposed theoretical framework. Additionally, we have
examined the stability using methods involving sound speed and the
adiabatic index. In recent years, the discipline of theoretical
physics has been greatly motivated in studying compact spheres.
These celestial entities provide considerable obstacles to our
understanding of basic physics and offer distinct possibilities for
studying extreme situations in the theoretical framework. Our study
is centered around the investigation of compact spheres in the
framework of $f(\textsc{Q},\textsc{T})$ theory with the objective of
exploring the enigmatic aspects of the cosmos. By examining compact
spheres in this theoretical framework, gravitational interactions
unveil on both galactic and cosmic levels. This offers valuable
understanding of how these components impact stellar structures.
Studying these objects in this theory allows us to investigate the
behavior of gravity under conditions of high curvature and density
as gravity in compact stars approaches its extreme limits. By
examining the behavior of these compact celestial objects, we
acquire significant knowledge about the properties of dense stellar
objects, enhancing our understanding of fundamental interactions in
the cosmos.

Metric coefficients show positively increasing behavior, ensuring a
smooth spacetime without any singularities (Figure \textbf{1}). The
central region of the considered charged spheres exhibit positive,
regular and maximal matter contents, implying a stable core that
reduces towards the boundary (Figure \textbf{2}). This feature
enhances the physical viability of the charged compact spheres. The
proposed stellar objects have a negative gradient of matter
contents, indicating a dense profile of the charged compact stellar
objects (Figure \textbf{3}). The diminishing anisotropy at the core
of charged spheres is a desirable characteristic for maintaining
their stability (Figure \textbf{4}). The positive behavior of all
energy constraints ensure that our proposed charged spheres are
viable (Figure \textbf{5}). The proposed charged star candidates are
viable as the EoS parameters range from 0 to 1 (Figure \textbf{6}).
The mass function increases with radial distance and vanishes at the
core limit (Figure \textbf{7}). Physically viable stellar objects
exist in modified $f(\textsc{Q},\textsc{T})$ theory, according to
the necessary viability conditions (Figure \textbf{8}). The cracking
condition is satisfied in the presence of modified terms (Figure
\textbf{9}). According to the TOV equation, it can be inferred that
the suggested charged stellar objects are in a state of equilibrium
(Figure \textbf{10}). Stellar objects are stable in this theory as
the stability limits are satisfied (Figures
\textbf{11}-\textbf{12}).

We have assessed the stability of our system in the presence of
modified terms. It has been noted that despite these modifications,
our system remains stable. Significantly, we notice that all
parameters attain their maximum levels in comparison to both GR and
other altered gravitational theories. In the context of
$f(\textsc{R},\textsc{T}^{2})$ theory, the compact stellar objects
are neither theoretically feasible nor stable at the center
\cite{25g}. Therefore, it is deduced from these results that all the
charged spheres exhibit both physical feasibility and stability at
their cores in this theory. Hence, our results indicate that this
theoretical framework support the existence of viable and stable
charged spheres.

%


\begin{thebibliography}{45}

\bibitem{1} Baade, W. and Zwicky, F.: Phys. Rev. \textbf{46}(1934)76.

\bibitem{2} Longair, M.S.: \emph{High Energy Astrophysics} (Cambridge Univeristy
Press, 1994).

\bibitem{3} Herrera, L., Santos, N.O.: Phys. Rep. \textbf{286}(1997)53.

\bibitem{4} Dev, K. and Gleiser, M.: Gen. Relativ. Gravit.
\textbf{39}(2002)1793.

\bibitem{5} Mak, M.K. and Harko, T.: Int. J. Mod. Phys. D
\textbf{13}(2004)156.

\bibitem{7} Rahaman, F. et al.: Eur. Phys. J. C \textbf{72}(2012)2071.

\bibitem{8} Hossein, S.K.M. et al.: Int. J. Mod. Phys. D \textbf{21}(2012)1250088.

\bibitem{9} Kalam, M. et al.: Eur. Phys. J. C \textbf{72}(2012)2248.

\bibitem{9a} Jeans, J.H.: Mon. Not. R. Astron. Sot. \textbf{82}(1922)122.

\bibitem{9b} Binney, J.: Ann. Rev. Astron. Astrophys. \textbf{20}(1982)399.

\bibitem{9c} Lemaitre, G.:. Ann. Sot. Sci. Bruxelles A \textbf{53}(1933)51.

\bibitem{9d} Bowers, R. and Liang, E.: Astrophys. J. \textbf{188}(1974)657.

\bibitem{9e} Bayin, S.: Phys. Rev. D \textbf{26}(1982)1262.

\bibitem{9f} Herrera, L. and Ponce de Leon, J.: J. Math. Phys. \textbf{26}(1985)2018.

\bibitem{9g} Herrera, L. and Ponce de Leon, J.: J. Math. Phys. \textbf{26}(1985)2847.

\bibitem{9h} Singh, K. and Bhamra, K.: Int. J. Theor. Phys. \textbf{29}(1990)1015.

\bibitem{9i} Gokhroo, M. and Mehra, A.: Gen. Rel. Grav. \textbf{26}(1994)75.

\bibitem{9j} Bondi, H.: Mon. Not. R. Astron. Sot. \textbf{259}(1992)365.

\bibitem{9k} Herrera, L. and Ponce de Leon, J.: J. Math. Phys. \textbf{26}(1985)2302.

\bibitem{9l} Herrera, L. and Varela, V.: Phys. Lett. A \textbf{189}(1994)11.

\bibitem{9m} Herrera, L., Ruggeri, G. and Witten, L.: Astrophys. J. \textbf{234}(1979)
1094.

\bibitem{9n} Chan, R., Herrera, L. and Santos, N.O.: Class. Quantum Grav. \textbf{9}(1992)
133.

\bibitem{9o} Herrera, L. and Santos, N.O.: Phys. Report \textbf{286}(1997)53.

\bibitem{9p} Chan, R., Herrera, L. and Santos, N.O.: Mon. Not. R. Astron. Sot. \textbf{265}
(1993)533.

\bibitem{9q} Herrera, L.: Phys. Rev. D \textbf{101}(2020)104024.

\bibitem{10} Weyl, H.S.: Preuss. Akad. Wiss. \textbf{1}(1918)465.

\bibitem{11} Jimenez, J.B., Heisenberg, I. and  Koivisto, L.T.: Phys. Rev
\textbf{98}(2018)044048.

\bibitem{12} Xu, Y. et al.: Eur. Phys. J. C \textbf{79}(2019)708.

\bibitem{4a} Cognola, G. et al.: Phys. Rev. D
\textbf{77}(2008)046009.

\bibitem{4b} Felice, A.D. and Tsujikawa S.R.: Living Rev. Relativ.
\textbf{13}(2010)161.

\bibitem{4c} Jawad, A. and Iqbal, A.: Int. J. Mod. Phys. D
\textbf{25}(2016)1650074.

\bibitem{4d} Jawad, A. and Rani, S.: Eur. Phys. J. C \textbf{76}(2016)704.

\bibitem{4e} Jawad, A. et al.: Astrophys. Space Sci.
\textbf{362}(2017)63.

\bibitem{5a} Sharif, M., Gul, M.Z.: Eur. Phys. J. Plus \textbf{133}(2018)345.

\bibitem{5b} Sharif, M., Gul, M.Z.: Int. J. Mod. Phys. D \textbf{28}(2019)1950054.

\bibitem{5c} Sharif, M., Gul, M.Z.: Chin. J. Phys. \textbf{57}(2019)329.

\bibitem{5d} Gul, M.Z. and Sharif, M.: New Astron. \textbf{106}(2024)102137.

\bibitem{5e} Sharif, M., Gul, M.Z.: Ann. Phys. \textbf{465}(2024)169674.

\bibitem{5f} Sharif, M., Gul, M.Z.: Phys. Scr. \textbf{99}(2024)065036.

\bibitem{5ff} Sharif, M., Gul, M.Z. and Hashim, I.: Phys. Dark Universe
\textbf{46}(2024)101606.

\bibitem{5g} Gul, M.Z., Sharif, M. and Hashim, I.: Phsys. Dark Universe \textbf{45}(2024)101537.

\bibitem{5h} Gul, M.Z. and Sharif, M.: Phys. Scr. \textbf{99}(2024)055036.

\bibitem{5i} Gul, M.Z. and Sharif, M.: Chin. J. Phys.
\textbf{88}(2024)388.

\bibitem{5ii} Jawad, A. et al.: Phys. Dark Universe
\textbf{46}(2024)101631.

\bibitem{5iii} Gul, M.Z., Sharif, M. and Kanwal, I.: New Astron. \textbf{109}(2024)102204.

\bibitem{5j} Jawad, A. et al.: Chin. J. Phys. \textbf{90}(2024)275.

\bibitem{15a} Xu, Y., Harko, T., Shahidi, S., and Liang, S.D.: Eur. Phys.
J. C \textbf{80}(2020)449.

\bibitem{15b} Arora, S. and Sahoo, P.K.: Phys. Scr. \textbf{95}(2020)095003.

\bibitem{15c} Bhattacharjee, S. and Sahoo, P.K.: Eur. Phys. J. C \textbf{80}(2020)289.

\bibitem{15d} Arora, S. et al.: Phys. Dark Universe \textbf{30}(2020)100664.

\bibitem{15e} Agrawal, A.S., Pati, L., Tripathy, S.K., and Mishra, B.:
Phys. Dark Universe \textbf{33}(2021)100863.

\bibitem{15f} Godani, N. and Samanta, G.C.: Int. J. Geom. Methods Mod. Phys. \textbf{18}(2021)2150134.

\bibitem{15g} Najera, A. and Fajardo, A.:
Phys. Dark Universe \textbf{34}(2021)100889.

\bibitem{15h} Arora, S., Santos, J.R.L. and Sahoo, P.K.: Phys. Dark Universe
\textbf{31}(2021)100790.

\bibitem{15i} Gul, M.Z., Sharif, M. Arooj, A.: Fortschr. Phys. \textbf{72}(2024)2300221.

\bibitem{15j} Gul, M.Z. et al.: Eur. Phys. J. C \textbf{84}(2024)775.

\bibitem{15k} Gul, M.Z., Sharif, M. Arooj, A.: Gen. Relativ. Gravit. \textbf{56}(2024)45.

\bibitem{15l} Gul, M.Z., Sharif, M. Arooj, A.:  Phys. Scr. \textbf{99}(2024)045006.

\bibitem{15m} Nan, G. et al.: Phsys. Dark Universe  \textbf{46}(2024)101635.

\bibitem{16a} Javed, F. et al.: Nucl. Phys. B \textbf{990}(2023)116180.

\bibitem{16b} Javed, F. et al.: Eur. Phys. J. C \textbf{83}(2023)1088.

\bibitem{16c} Javed, F. and Lin, J.: Chin. J. Phys. \textbf{88}(2024)786.

\bibitem{16d} Mustafa, G. et al.: Phys. Dark Universe \textbf{30}(2020)
100652.

\bibitem{23} Nashed, G.G. and Capozziello, S.: Eur. Phys. J. C \textbf{81}(2021)481.

\bibitem{24} Kumar, J., Singh, H.D. and Prasad, A.K.: Phys. Dark
Universe \textbf{34}(2021)100880.

\bibitem{25} Dey, S., Chanda, A. and Paul, B.C.: Eur. Phys. J. Plus \textbf{136}(2021)228.

\bibitem{25a} Sharif, M. and Gul, M.Z.: Chin. J. Phys. \textbf{71}(2021)365.

\bibitem{25b} Sharif, M. and Gul, M.Z.: Universe \textbf{96}(2021)154.

\bibitem{25c} Sharif, M. and Gul, M.Z.: Int. J. Mod. Phys. A \textbf{36}(2021)2150004.

\bibitem{25d} Sharif, M. and Gul, M.Z.: Adv. Astron. \textbf{2021}(2021)6663502.

\bibitem{25e} Sharif, M. and Gul, M.Z.:  Int. J. Geom. Methods Mod. Phys. \textbf{19}(2022)2250012.

\bibitem{25f} Sharif, M. and Gul, M.Z.: Mod. Phys. Lett. A \textbf{19}(2022)2250005.

\bibitem{25g} Sharif, M. and Gul, M.Z.: Gen. Relative. Gravit. \textbf{55}(2023)10.

\bibitem{25h} Sharif, M. and Gul, M.Z.: Fortschr. Phys. \textbf{71}(2023)2200184.

\bibitem{25i} Sharif, M. and Gul, M.Z.: Phys. Scr. \textbf{98}(2023)035030.

\bibitem{25j} Sharif, M. and Gul, M.Z.: Pramana-J. Phys. \textbf{97}(2023)122.

\bibitem{26} Majeed, A., Abbas, G. and Shahzad, M.R.:
New Astron. \textbf{102}(2023)102039.

\bibitem{27a} Adeel, M. et al.: Mod. Phys. Lett. A \textbf{38}(2023)2350152.

\bibitem{27b} Gul, M.Z. et al.: Eur. Phys. J. C \textbf{84}(2024)8.

\bibitem{27c} Rani, S. et al.: Int. J. Geom. Methods Mod. Phys.
\textbf{21}(2024)2450033.

\bibitem{28a} Sharif, M. and Gul, M.Z.: Ann. Phys.
\textbf{465}(2024)169674.

\bibitem{28b} Sharif, M. and Gul, M.Z.: Phys. Scr. \textbf{99}(2024)065036.

\bibitem{28c} Sharif, M., Shakeel, M. and Gul, M.Z.:  New Astron. \textbf{108}(2024)102179.

\bibitem{28} Bhar, P. and Pretel, J.M.: Phys. Dark Universe
\textbf{42}(2023)101322.

\bibitem{29} Ilyas, M. and Ahmad, D.: Chin. J. Phys.
\textbf{88}(2024)901.

\bibitem{30} Rej, P. and Bhar, P.: New Astron.
\textbf{105}(2024)102113.

\bibitem{31} Das, K.P. et al.:
Phys. Dark Universe \textbf{43}(2024)101398.

\bibitem{34} Tolman, R.C.: Phys. Rev. D \textbf{55}(1939)364.

\bibitem{35} Ozel, F., Guver, T. and Psaltis, D.: Astrophys. J. \textbf{693}(2009)1775.

\bibitem{36} Elebert, P. et al.: Mon. Not. R. Astron. Soc. \textbf{395}(2009)884.

\bibitem{37} Ozel, F., Guver, T., Cabrera-Lavers, A. and Wroblewski, P.: Astrophys. J. \textbf{712}(2010)964.

\bibitem{38} Guver, T. et al.: Astrophys. J. \textbf{719}(2010)1807.

\bibitem{39} Demorest, P.B.: Nature \textbf{467}(2010)1081.

\bibitem{40} Rawls, M.L. et al.: Astrophys. J. \textbf{730}(2011)25.

\bibitem{41} Freire, P.C.C. et al.: Mon. Not. R. Astron. Soc. \textbf{412}(2011)2763.

\bibitem{42} Singh, K.N. et al.: Eur. Phys. J. A \textbf{53}(2017)21.

\bibitem{44} Buchdahl, A.H.: Phys. Rev. D \textbf{116}(1959)1027.

\bibitem{45} Ivanov, B.V.: Phys. Rev. D \textbf{65}(2002)104011.

\bibitem{45a} Herrera, L. and Di Prisco, A.: Phys. Rev. D
\textbf{109}(2024)064071.

\bibitem{45b} Herrera, L.: Phys. Lett. A \textbf{165}(1992)206.

\bibitem{46} Tolman, R.C.: Phys. Rev. \textbf{55}(1939)364; Oppenheimer,
J.R. and Volkoff, G.M.: Phys. Rev. \textbf{55}(1939)374.

\bibitem{47} Abreu, H. et al.: Class. Quantum Grav.
\textbf{24}(2007)4631.

\bibitem{48} Chandrasekhar, S.: Mon. Not. R. Astron. Soc. \textbf{140}(1964)417.

\bibitem{49} Bondi, H.: Proc. R. Soc. London A \textbf{281}(1964)39.

\bibitem{50} Chan, R. et al.: Mon. Not. R. Astron. Soc. \textbf{265}(1993)533.

\bibitem{51} Chan, R. et al.: Class. Quantum Grav. \textbf{9}(1992)133.

\end{thebibliography}
\end{document}